\documentstyle[twocolumn,eqsecnum,prb,aps]{revtex}

\begin{document}   
 
\draft  

\title{Electromagnetic characteristics and effective gauge theory \\
of double-layer quantum Hall systems}
 \author{K. Shizuya}
  \address{Yukawa Institute for Theoretical Physics\\
 Kyoto University,~Kyoto 606-8502,~Japan }

\maketitle

\begin{abstract} 
The electromagnetic characteristics of double-layer quantum Hall
systems are studied, with projection to the lowest Landau level
taken into account and intra-Landau-level collective excitations
treated in the single-mode approximation. It is pointed out
that dipole-active excitations, both elementary and
collective, govern the long-wavelength features of quantum Hall
systems. In particular, the presence of the dipole-active
interlayer  out-of-phase collective excitations, inherent to
double-layer systems, modifies the leading $O({\bf k})$ and
$O({\bf k}^{2})$ long-wavelength characteristics (i.e., the transport
properties and characteristic scale) of the double-layer
quantum Hall states substantially.
We apply bosonization techniques and construct from such
electromagnetic characteristics an effective theory, which
consists of three vector fields representing the three
dipole-active modes, one interlayer collective mode and two
inter-Landau-level cyclotron modes.
This effective theory properly incorporates the spectrum
of collective excitations on the right scale of the Coulomb
energy and, in addition, accommodates the favorable
transport properties of the standard Chern-Simons theories.
\end{abstract}


\section{Introduction}

Laughlin's variational wave functions~\cite{L} gave
impetus~\cite{GM} to descriptions of the fractional quantum
Hall effect~\cite{TSG,Rev} (FQHE) in terms of electron--flux
composites.  The Chern-Simons theories realize the 
composite-boson~\cite{ZHK,LZ,Z} and composite-fermion~\cite{BW,LF,HLR}
pictures of the FQHE and have been successful in describing various aspects
of the fractional quantum Hall (FQH) states.

An alternative approach,~\cite{KSbos} that makes use of projection
to Landau levels and bosonization as basic
tools, has recently been developed to study the long-wavelength
features of single-layer quantum Hall systems.
There an effective vector-field theory is constructed,
via bosonization, from the electromagnetic response of
incompressible  states.
It refers to neither the composite-boson nor composite-fermion
picture, but properly reproduces the results of the Chern-Simons
(CS) theories at long wavelengths, thus revealing the universal
long-wavelength characteristics of the incompressible FQH states.

For double-layer systems the quantum Hall effect
exhibits a rich variety of patterns and physics, as observed
experimentally~\cite{SES} and discussed
theoretically.~\cite{BIH,Chak,MPB,RR,Yang,MZ,Moon,WZdlayer,EI,LFdlayer}
The Chern-Simons theoretic approaches, however, encounter some subtle
problem, when generalized to double-layer systems.
They differ significantly~\cite{MZ} in the
collective-excitation spectrum from a general magneto-roton theory of
collective excitations based on the single-mode approximation (SMA),
developed by Girvin, MacDonald and Platzman.~\cite{GMP}

The purpose of this paper is to extend the previous (projection
+ bosonization) approach to double-layer quantum Hall systems and
to clarify the relation between the microscopic SMA theory and
the CS theories.  
Double-layer systems generally support dipole-active intra-Landau-level
collective modes,~\cite{RR,MZ} in sharp contrast to single-layer systems
where only the cyclotron mode remains dipole active.~\cite{GMP}
[A dipole-active mode is the one whose spectral weight behaves like
$O({\bf k}^{2})$ in the long-wavelength limit ${\bf k} \rightarrow 0$. 
%, directly measurable in experiments,
It is, as such, sensitive to long-wavelength probes.]
We study the electromagnetic response of a double-layer system
in the absence of interlayer tunneling, with the collective
excitations taken into account in the SMA. It is shown that
the presence of the dipole-active interlayer collective excitations 
modifies the $O({\bf k})$ and $O({\bf k}^{2})$ long-wavelength
response (i.e., the transport properties and characteristic scale) of the
system substantially. The effective theory of the FQHE, reconstructed from
this response,  consists of three vector fields that represent the three
dipole-active modes residing in the system. It properly embodies the SMA
spectrum of collective excitations, along with the favorable transport
properties of the CS theories.

In Sec.~II we study the electromagnetic response of a single-layer
Hall-electron system and present a general formalism for achieving
projection onto the true Landau levels by a suitable unitary or
$W_{\infty}$ transformation. 
In Sec.~III we examine the electromagnetic characteristics of a
double-layer system and, in Sec.~IV, construct an effective gauge theory
of the FQHE.
Section~V is devoted to a summary and discussion.

\section{Electromagnetic response of Hall electrons}

Consider electrons confined to a plane with a perpendicular
magnetic field $B_{z}=B>0$. Our task in this section is to
study how they respond to weak external potentials
$A_{\mu}({\bf x},t)=(A_{0}({\bf x},t), {\bf A}({\bf x},t))$, 
with the action
\begin{eqnarray}
S_{1} &=& \int dt d^{2}{\bf x}\, \psi^{\dag}({\bf x},t)\,
(i\partial_{t}- {\cal H})\, \psi ({\bf x},t),
\label{Selectron}\\
{\cal H} &=& {1\over{2M}} \left({\bf p}+e{\bf A}^{\!B}({\bf x})+e{\bf
A}({\bf x},t)\right)^{2} +eA_{0}({\bf x},t) ,
\label{Hone}
\end{eqnarray}
where %the Landau-gauge potential
${\bf A}^{\!B}={1\over{2}}B\,(-y,0)$
supplies a uniform magnetic field $B$. 
The Coulomb interaction will be included later. 
[We suppose that $\mu$ runs over $(t,x,y)$ or $(0,1,2)$, and denote 
${\bf A}=(A_{1},A_{2})$, etc.]
% Remember that our $A_{0}$ equals minus the
%conventional scalar potential.]
For conciseness we shall write
$x=(t,{\bf x}),  d^{3}x = dt\,d^{2}{\bf x},
A_{\mu}(x) = A_{\mu}({\bf x},t)$, etc.,  when no confusion arises; 
the electric charge $e>0$ is also suppressed by rescaling
$eA_{\mu}\rightarrow A_{\mu}$ in what follows.

The eigenstates of a free orbiting electron are Landau levels
$|N\rangle = |n, y_{0}\rangle$ of energy 
$\omega_{c} (n+{\scriptstyle {1\over2}})$, labeled by integers
$n = 0,1,2,\cdots$, and $y_{0}=\ell^{2}\,p_{x}$, where
$\omega_{c} \equiv eB/M$ and ${\ell}\equiv 1/\sqrt{eB}$.
This level structure is modified in the presence of  $A_{\mu}(x)$, and
the effect of level mixing it causes is calculated by diagonalizing the
Hamiltonian with respect to the true Landau levels $\{n\}$.

The actual calculation is best done in $N$ space, where 
the Hamiltonian $\langle n,y_{0}|{\cal H}|n',y_{0}'\rangle 
= \langle y_{0}|\hat{\cal H}_{nn'}|y_{0}'\rangle$ becomes an
infinite-dimensional matrix in level indices and an operator in $y_{0}$ 
and its conjugate $x_{0}\equiv i\ell^{2} \partial/\partial y_{0}$;   
${\bf r}= (x_{0}, y_{0})$ represents the center coordinate of an
orbiting electron with uncertainty $[r_{1}, r_{2}]=i\ell^{2}$. 
With a unitary transformation
$\psi ({\bf x},t)=\sum_{N} \langle {\bf x}|N\rangle\, \psi_{n}(y_{0},t)$, 
the one-body action is rewritten as~\cite{KSw}
\begin{eqnarray}
 S_{1} &=& \int\! dt\, d y_{0}\! \sum_{m,n=0}^{\infty}\!
\psi^{\dag}_{m}(y_{0},t) \left(i\delta_{m n} \partial_{t}
-{\hat{\cal H}}_{m n} \right) \psi_{n}(y_{0},t),  \nonumber \\
\hat{\cal H}&=& \omega_{c} (Z^{\dag} Z + {\scriptstyle {1\over2}}) +
U(\hat{\bf x},t),
\label{Hczzdag}
\end{eqnarray}
where
\begin{eqnarray}
U(x)&=&i\ell \omega_{c} \{Z^{\dag}A(x)-A^{\dag}(x)Z\}
 +\ell^{2}\omega_{c} A^{\dag}(x)A(x)
\nonumber\\ &&
+ A_{0}(x)+ {1\over{2M}}\, A_{12}(x)
\label{Udef} 
\end{eqnarray}
summarizes the electromagnetic coupling and
\begin{eqnarray}
 A &=& (A_{2} +iA_{1})/\sqrt{2},\ A^{\dag} = (A_{2}
-iA_{1})/\sqrt{2} ;
%\nonumber\\
% A_{12}& =&\partial_{1}A_{2}-\partial_{2}A_{1}.
\end{eqnarray} 
$A_{12} =\partial_{1}A_{2}-\partial_{2}A_{1}$.  Note that in $N$ space the
coordinate operators ${\bf x}= (x_{1},x_{2})$ take the matrix form 
\begin{eqnarray}
(\hat{x}_{i})_{nn'}=(X_{i})_{nn'}+r_{i}\delta_{nn'},
\label{xhat}
\end{eqnarray}
where $X_{i}$ are the hermitian matrices of a harmonic oscillator with 
$[X_{1}, X_{2}]=-i\ell^{2}$; in the above, we have defined 
$Z=(X_{2} +iX_{1})/(\sqrt{2} \ell)$ so that
$Z_{mn}=\sqrt{n}\,\delta_{m,n-1}$ and $[Z, Z^{\dag}]=1$. 
In this section we use the Landau gauge but all the manipulations are
carried over to the symmetric gauge~\cite{GMP,GJach} as well.

The one-body action $S_{1}$ is made diagonal in level indices by a suitable
$U(\infty)$ or $W_{\infty}$ transformation $G$ of the form 
\begin{equation}
\psi^{G}_{m}(y_{0},t)= \sum_{n=0}^{\infty}
G_{mn}(x_{0},y_{0},t)\,\psi_{n}(y_{0},t),
\end{equation}
under which  $U$ undergoes the transformation 
\begin{equation}
U^{G}=G U G^{-1} +\omega_{c} [G, Z^{\dag}Z]G^{-1} -iG \partial_{t}G^{-1}.
\label{VG} 
\end{equation}
A general program of projection onto the true Landau levels along this
line was laid out earlier.~\cite{KSw}  Here we complete it by presenting an $O(U^{2})$
expression, exact to all powers of derivatives, and how to handle the
Coulomb interaction.

For diagonalization let us expand $U$ in the $U(\infty)$
basis $\{\Gamma_{su}\equiv (Z^{\dag})^{s}Z^{u}/(s!u!)\}$, 
\begin{equation}
U(\hat{\bf x},t) =\sum_{s,u=0}^{\infty}U_{su}({\bf r},t)\, \Gamma_{su},
\label{Uinfinity}
\end{equation}
where
${\bf r}=(r_{1},r_{2})$.  Likewise, we write $G=\exp[i\ell\,\eta]$ and
$\eta=\sum_{s,u=0}^{\infty}\eta_{su}({\bf r},t)\,\Gamma_{su}$.
With the choice
\begin{equation}
i\eta_{su} = [(s-u)\omega_{c} -i\partial_{t}]^{-1}U_{su}\ \ \ \
(s\not=u),
\label{etaone}
\end{equation} 
$U^{G}= \sum_{s,u=0}^{\infty}(U^{G})_{su}\Gamma_{su}$ is
made diagonal to $O(U^{2})$.
In particular, for the lowest Landau level
\begin{equation}
 (U^{G})_{00} = U_{00} -\sum_{s=1}^{\infty} {1\over{s!}}\, 
U_{0s}\ {1\over{s\omega_{c} - i\partial_{t}}} \ U_{s0} +\cdots,
\label{VGdiag} 
\end{equation}
apart from an (unimportant) total divergence %terms 
$\propto \partial_{t}(\cdots)$.
In what follows we focus on the lowest Landau level
(of practical interest) under a strong magnetic field. 
%For notational simplicity we shall from now on set 
For conciseness we set the magnetic length $\ell \rightarrow 1$ below.

Let us write out $U_{su}$.  To this end define the
field $A(\hat{\bf x},t)$ through the Fourier transform 
$A(\hat{\bf x},t)=\sum_{\bf p} A[{\bf p},t] e^{i{\bf p\cdot \hat{x}} }$ 
and normal-order it with respect to $Z^{\dag}$ and $Z$,
\begin{eqnarray}
A(\hat{\bf x},t) &=&\sum_{\bf p}A[{\bf p},t]\,F({\bf p})\,
e^{-{1\over{4}}\, {\bf p}^{2}}\, e^{i{\bf p\cdot r}}, \\
F({\bf p}) &\equiv& :e^{i{\bf p\cdot X}}:\,
= e^{{i\over{\sqrt{2}}}\,p Z^{\dag}}\,
e^{{i\over{\sqrt{2}}}\,p^{\dag} Z},
\label{Fp}
\end{eqnarray}
where $p=p_{2}+ ip_{1}$ and $p^{\dag}=p_{2}- ip_{1}$.
One can thus write
\begin{eqnarray}
A(\hat{\bf x},t)&=&\sum_{s,u=0}^{\infty} \Gamma_{su}\, 
\bar{\partial}^{s} \partial^{u}
e^{{1\over{2}}\bar{\partial}\partial}\,A({\bf r},t),
\label{Ahatx}
\end{eqnarray}
with
$\bar{\partial}={1\over{\sqrt{2}}}(\partial_{r_{2}} +i\partial_{r_{1}})$
and $\partial={1\over{\sqrt{2}}}(\partial_{r_{2}} -i\partial_{r_{1}})$
acting on $A({\bf r},t)$.
One may regard 
$\bar{\partial}^{s}
\partial^{u} e^{{1\over{2}}\bar{\partial}\partial}\,A({\bf r},t)$
as representing the Fourier transform
$(ip/\sqrt{2})^{s} (i p^{\dag}/\sqrt{2})^{u}e^{-{1\over{4}}\, 
{\bf p}^{2}}\, A[{\bf p},t]$.
Substitution of Eq.~(\ref{Ahatx}) into $U$ yields
\begin{eqnarray}
U_{00}&=& A_{0}^{[{\bf r}]} + {1\over{2M}}\,A_{12}^{[{\bf r}]} +
\omega_{c} (A^{\dag}A)^{[{\bf r}]}\equiv \chi^{[{\bf r}]}, \nonumber\\
U_{s0}&=&\bar{\partial}^{s-1} ( i s \omega_{c} A^{[{\bf r}]}  +
\bar{\partial} \chi^{[{\bf r}]}) \ \ \ \ (s\ge 1),\nonumber\\ 
U_{0s}&=& \partial^{s-1} ( -i s \omega_{c} 
\bar{A}^{[{\bf r}]}  + \partial \chi^{[{\bf r}]}),
\end{eqnarray}
where 
$A^{[{\bf r}]} \equiv e^{{1\over{2}}\bar{\partial}\partial} A({\bf r},t), 
A_{12}^{[{\bf r}]} \equiv e^{{1\over{2}}\bar{\partial}\partial}
A_{12}({\bf r},t)$, etc.

 Some care is needed in handling operators involving ${\bf r}$ with
$[r_{1},r_{2}]=i$. 
% alculating $(U^{G})_{00}$ since
%$U_{su}$ contain noncommuting operators ${\bf r}$ with $[r_{1},r_{2}]=i$. 
Note that
$T_{\bf p}  = e^{-{1\over{4}}\,{\bf p}^{2}}\, e^{i{\bf p\cdot r}}$
obeys the multiplication law 
\begin{eqnarray}
T_{\bf p}\, T_{\bf k}
= e^{{1\over{2}}\, p^{\dag}\,k}\, T_{\bf p+k} 
= e^{{1\over{2}}\,({\bf p\cdot k}-i\, {\bf p}\times {\bf k})}\,
T_{\bf p+k},
\end{eqnarray}
where ${\bf p}\times{\bf k} \equiv
\epsilon^{ij}p_{i}k_{j}=p_{1}k_{2}- p_{2}k_{1}$ 
with $\epsilon^{12} =1$.
One can therefore write an operator product  
$A^{[{\bf r}]}B^{[{\bf r}]}$ as
\begin{eqnarray}
A^{[{\bf r}]}B^{[{\bf r}]}
&=&\sum_{\bf p}T_{\bf p}\, (A*B)_{\bf p},\\
(A*B)_{\bf p} &=&\sum_{\bf k}
e^{{1\over{2}}\, (p^{\dag}-k^{\dag}) k}\,
A[{\bf p-k},t] B[{\bf k},t].
\end{eqnarray}
Thus, back in ordinary (commuting) ${\bf x}$ space, 
\begin{eqnarray}
A*B &=& e^{- \partial \bar{\partial}'}\, A({\bf x})
B({\bf x'}) |_{ {\bf x'} \rightarrow {\bf x}} 
\nonumber\\
&=&  A({\bf x})  B({\bf x})-
\partial A({\bf x})\, \bar{\partial} B({\bf x})  +\cdots.
\end{eqnarray}

With this rule one can translate operator products into ordinary functions 
and go through the calculation.  Let us write 
$U^{G}_{00}({\bf r},t) = \sum_{\bf p} e^{i{\bf p\cdot r}}
e^{-{1\over{4}}\, {\bf p}^{2}} {\cal U}^{G}_{00}[{\bf p},t]$ with
${\cal U}^{G}_{00}[{\bf p},t]=  A_{0}[{\bf p},t]
+ {1\over{2M}}A_{12}[{\bf p},t] + {\cal U}_{2}[{\bf p},t]$.  
The ${\cal U}_{2}[{\bf p},t]$ denotes the contribution
quadratic in $A_{\mu}$.
We here  quote only the ${\bf p}=0$ component 
${\cal U}_{2}[{\bf p}=0,t]= \int d^{2}{\bf x}\, {\cal U}_{2}(x)$  with
\begin{eqnarray}
{\cal U}_{2}(x) 
&=& - {1\over{2\omega_{c}}} \Big(A_{k0}\! +\!{1\over{2M}}\partial_{k}
A_{12}\Big)\,D \Big(A_{k0}\! + \!{1\over{2M}}\partial_{k} A_{12}\Big)
\nonumber\\ && 
+{1\over{2}} A_{\mu} D'\epsilon^{\mu \nu \rho}\partial_{\nu}A_{\rho}
+{1\over{2M}} \, A_{12} D'\, A_{12}  
\label{Utwo}
\end{eqnarray}
in compact notation.
 Here the correlation functions
$D=\sum_{n=1}^{\infty} D_{n}$  and
$D'=\sum_{n=1}^{\infty} nD_{n}$ are written in terms of
\begin{eqnarray}
D_{n}  ={1\over{(n-1)!}}\,{\omega_{c}^{2}\over{(n\omega_{c})^{2}
\!+\! \partial_{t}^{2}}}\, e^{{1\over{2}}\nabla^{2}}
(-{\scriptstyle {1\over{2}}} \nabla^{2})^{n-1},
\end{eqnarray}
with $\nabla^{2} \equiv \partial_{k}\partial_{k}$;
$A_{\mu \nu}= \partial_{\mu}A_{\nu}- \partial_{\nu}A_{\mu}$, and
$\epsilon^{\mu\nu\rho}$ is a totally-antisymmetric
tensor with $\epsilon^{012} =1$.

The electromagnetic coupling projected to the lowest Landau level is 
now written as
\begin{eqnarray}
&&\bar{H}^{\rm em}\!=\! \sum_{\bf p}\! \Big\{
A_{0}[{\bf p},t] +\! {1\over{2M}}A_{12}[{\bf p},t]
+ {\cal U}_{2}[{\bf p},t] \Big\}\, \rho_{G}^{(00)}[{\bf -p},t],
\nonumber\\
&&\\
 &&\rho_{G}^{(00)}[{\bf p},t] \equiv
\int dy_{0}\ {\psi^{G}_{0}}^{\dag}(y_{0},t)\,
e^{-{1\over{4}}\, {\bf p}^{2}}\, e^{-i{\bf p\cdot r}}\,
\psi^{G}_{0}(y_{0},t).\nonumber\\
&&
\end{eqnarray}
This $\rho_{G}^{(00)}[{\bf p},t]$ is the basic charge operator we use.
It differs slightly from the charge operator projected to the lowest
Landau level.  Indeed, expressing 
$\rho [{\bf p},t]=\int d^{2}x\, e^{-i{\bf p\cdot x}}\,
\psi^{\dag}(x)\psi(x)$ 
in terms of $\psi^{G}_{n}$ yields 
\begin{eqnarray}
\rho[{\bf p},t]
&=&  \int dy_{0}\, {\psi^{G}_{n}}^{\dag}(y_{0},t)\,
W_{nn'}\psi^{G}_{n'} (y_{0},t) ,\nonumber\\
W_{nn'}&=& G_{nm}\,F_{m m'}({\bf -p})\,
e^{-{1\over{4}}\,{\bf p}^{2}}\, e^{-i{\bf p\cdot r}}\, (G^{-1})_{m'n'},
\label{rhopt}
\end{eqnarray}
in obvious notation; 
$F_{mm'}({\bf -p})$ stands for the matrix % defined in Eq.
~(\ref{Fp}).
Let us extract the lowest-Landau-level component
$\int dy_{0}\, \psi^{G\dag}_{0}\,W_{00}\, \psi_{0}^{G}$ and denote
it as  $\rho^{(00)}[{\bf p},t] =  \rho_{G}^{(00)}[{\bf p},t] +
\triangle {\rho_{G}^{(00)}}[{\bf p},t]$.  
The deviation to $O(U)$ reads
\begin{eqnarray}
\triangle \rho_{G}^{(00)}[{\bf p},t]
&=& \sum_{\bf k} 
 \rho^{(00)}_{G}[{\bf p\! -\!k},t]  \ u[{\bf p,k},t],
\label{trirhoG}
\end{eqnarray}
with
\begin{eqnarray}
u[{\bf p,k},t] &=& i e^{{1\over{4}}\,{\bf k}^{2}}\,
\sum_{s=1} {1\over{s!}}\,\Big\{ e^{-{1\over{2}}k^{\dag} p}\,
 \Big( {{-ip\over{\sqrt{2}}}} \Big)^{s}\, \eta_{0s}[{\bf k},t]
\nonumber\\ &&
- e^{-{1\over{2}} p^{\dag} k}\,
\Big( {{-i p^{\dag}\over{\sqrt{2}}}} \Big)^{s}\,
\eta_{s0}[{\bf k},t]\Big\} \nonumber\\
&=& i p_{j}\Big\{\epsilon^{jk} A_{k}[{\bf k},t] 
+ O(A_{i0}) \Big\} + O({\bf k}).
\label{uzerozero}
\end{eqnarray}
Here we have retained only terms with no derivatives ${\bf k}$ or
$\partial_{t}$ acting on $A_{\mu}[{\bf k},t]$, the portion needed later. 
As seen from Eq.~(\ref{rhopt}), $\rho[{\bf p},t]$ is not invariant
under $W_{\infty}$ transformations $G$, except for the total charge 
$\rho[{\bf p}=0,t]$. This explains why 
$\rho^{(00)}[{\bf p},t]$ differs from $\rho_{G}^{(00)}[{\bf p},t]$.

Let us next consider the Coulomb interaction, which is a functional of
the density $\rho (x)=\psi^{\dag}(x)\psi(x)$ or
the deviation $\delta \rho (x)=\rho (x)-\rho_{\rm av}$
from the average electron density $\rho_{\rm av}$.
Passing to the Fourier space and setting 
$\rho [{\bf p},t]\rightarrow \rho^{(00)}[{\bf p},t]$ yields the Coulomb
interaction projected to the lowest Landau level
\begin{eqnarray}
\bar{H}^{\rm Coul} &=& {1\over{2}}\, \sum_{\bf p} V[{\bf p}]\,
\delta \bar{\rho}_{\bf -p}\, \delta \bar{\rho}_{\bf p}
+ \triangle H^{\rm Coul},\\
\triangle H^{\rm Coul} &=&  {1\over{2}}\sum_{\bf p} V[{\bf p}]\, 
\sum_{\bf k} u[{\bf p,k},t]\, \{ \bar{\rho}_{\bf -p},
\bar{\rho}_{\bf p - k} \},
\label{triHC}
\end{eqnarray}
where we have denoted 
$\bar{\rho}_{\bf p} \equiv \rho_{G}^{(00)}[{\bf p},t]$ and
$\delta \bar{\rho}_{\bf p} \equiv \bar{\rho}_{\bf p} 
-\rho_{\rm av}\, (2\pi)^{2}\delta^{2} ({\bf p})$, with obvious
$t$ dependence suppressed; $V[{\bf p}] =V[-{\bf p}]$ denotes the Coulomb
potential in momentum space. In Eq.~(\ref{triHC}) we have set 
$\delta \bar{\rho}_{\bf -p}\rightarrow \bar{\rho}_{\bf -p}$ 
because the $\rho_{\rm av}$ term does not affect the
electromagnetic response as long as 
$V[{\bf 0}] <\infty $ or even for $V[{\bf p}] \sim 1/|{\bf p}|$. 
The $(W_{\infty}$-breaking) correction $\triangle H^{\rm Coul}$ has an
important consequence, as we shall see later.

The dynamics within the lowest Landau level is now governed by the
Hamiltonian $\bar{H}= \bar{H}^{\rm Coul} +\bar{H}^{\rm em}$.
[We have suppressed the quenched kinetic energy term
${1\over{2}}\,\omega_{c}$.]
Suppose now that an incompressible many-body state $|G\rangle$ of
uniform density $\rho_{\rm av}= \nu/(2\pi\ell^{2})$ is formed
within the lowest Landau level $(\nu < 1)$ via the Coulomb interaction
(for $A_{\mu}=0$).
Then its response to weak electromagnetic potentials
$A_{\mu}(x)$ is described by the rest of terms in $\bar{H}$.
In particular, setting $\langle G| \bar{\rho}_{\bf -p} |G
\rangle =\rho_{\rm av}\, (2\pi)^{2}\delta^{2} ({\bf p})$
in $\bar{H}^{\rm em}$ one obtains the effective action to $O(A^{2})$:
\begin{eqnarray}
S^{\rm em}= -\rho_{\rm av} \int d^{3}x \Big[ A_{0}(x) + 
{\cal U}_{2}(x) \Big] ,
\label{Sem}
\end{eqnarray}
which is manifestly gauge invariant.
The ${\cal U}_{2}(x)$ summarizes the effect of electromagnetic
inter-Landau-level mixing and agrees~\cite{fnutwo} with the result of
a direct perturbative calculation~\cite{LF} (for $\nu =1$). 
Note that this response is determined by the charge density 
$\rho_{\rm av}$ alone without knowing further details of 
the state $|G \rangle$.

The electromagnetic interaction in $\bar{H}$ also gives rise to
intra-Landau-level transitions.  The inter-level
cyclotron mode, however, saturates the oscillator-strength sum rule in
accordance with Kohn's theorem~\cite{Kohn} and, as a result, the
intra-Landau-level excitations are only
dipole-inactive~\cite{GMP}  (i.e., the response vanishes faster than
${\bf k}^{2}$ for ${\bf k} \rightarrow 0$); this implies, in
particular, that the quantum Hall states show universal $O({\bf k})$
and $O({\bf k}^{2})$ long-wavelength electromagnetic characteristics
determined by $S^{\rm em}$ above.
The situation changes drastically for double-layer (or multi-layer)
systems, which we discuss in the next section.

\section{Double-layer systems}

In this section we study the electromagnetic response of
double-layer quantum Hall systems.
Consider a double-layer system in the absence of interlayer tunneling,
with  average electron densities 
$\rho_{\rm av}^{\alpha}=(\rho_{\rm av}^{1},\rho_{\rm av}^{2})$ in the
upper $(\alpha =1)$ and lower $(\alpha =2)$ layers.  
The electron fields $\psi^{\alpha}$ in the two layers are taken to be
coupled through the Coulomb interaction 
\begin{eqnarray}
H^{\rm Coul}&=& {1\over{2}} \sum_{\bf p}\,
\delta \rho^{\alpha}[{\bf -p},t]\,
V_{\alpha\beta}[{\bf p}]\,  \delta \rho^{\beta}[{\bf p},t],
\end{eqnarray}
with $\delta \rho^{\alpha}[{\bf p},t]= \rho^{\alpha}[{\bf p},t]
- \rho_{\rm av}^{\alpha}\, (2\pi)^{2}\delta^{2} ({\bf p})$ and
$\rho^{\alpha}(x)={\psi^{\alpha}}^{\dag}(x)\psi^{\alpha}(x)$.
Here $V_{11}[{\bf p}]=V_{22}[{\bf p}]$ and 
$V_{12}[{\bf p}] =V_{21}[{\bf p}]$ are the intralayer and interlayer
potentials, respectively; 
summations over repeated layer-indices $\alpha=1,2$ are understood
from now on.

The system is placed in a common strong perpendicular magnetic field
$B$.  To probe each layer let us couple two weak fields 
$A_{\mu}^{\alpha}= (A_{\mu}^{1},A_{\mu}^{2})$ to the two layers separately,
and carry out projection onto the Landau levels for each layer. Then the
dynamics within the lowest Landau level is governed by the Hamiltonian
\begin{eqnarray}
\bar{H}&=&{1\over{2}}\, \sum_{\bf p} V_{\alpha \beta}[{\bf p}]\,
\delta \bar{\rho}^{\alpha}_{\bf- p}\,\delta \bar{\rho}^{\beta}_{\bf p}
\nonumber\\ && 
+ \sum_{\rm p}\Big(A_{0}^{\alpha}[{\bf p}] +
{1\over{2M}}\,A_{12}^{\alpha}[{\bf p}]\Big)\,
\bar{\rho}^{\alpha}_{\bf -p}
+ \triangle H^{\rm Coul},
\label{HDL}
\end{eqnarray}
where we have denoted 
$\bar{\rho}^{\alpha}_{\bf p}\equiv \rho_{G}^{\alpha (00)}[{\bf p},t]$,  
$A_{0}^{\alpha}[{\bf p},t] \rightarrow A_{0}^{\alpha}[{\bf p}]$, etc., 
for short. [For conciseness we shall suppress such obvious time dependence
in what follows; confusion may not arise since we mainly handle quantities
at equal times or at a fixed time.] The response due to the
cyclotron modes, including the $\rho^{\alpha}_{\rm av}A_{0}^{\alpha}$
coupling and isolated from $\bar{H}$, reads
\begin{eqnarray}
S^{\rm cycl}= -\int d^{3}x\, \rho^{\alpha}_{\rm av} \,
\Big[ A^{\alpha}_{0}(x) +{\cal U}_{2}^{\alpha}(x) \Big],
\label{Scycl}
\end{eqnarray}
where ${\cal U}_{2}^{\alpha}(x)$ stand for ${\cal U}_{2}(x)$
in Eq.~(\ref{Utwo}) with $A_{\mu} \rightarrow A_{\mu}^{\alpha}$.

It is convenient to use, instead of $\bar{\rho}^{\alpha}_{\bf p}$,
\begin{eqnarray}
\bar{\rho}_{\bf p}= \bar{\rho}^{1}_{\bf p} + \bar{\rho}^{2}_{\bf p}, \
\bar{d}_{\bf p}= \bar{\rho}^{1}_{\bf p} - \bar{\rho}^{2}_{\bf p},
\end{eqnarray}
and write the Coulomb interaction as
\begin{eqnarray}
\bar{H}^{\rm Coul}
=&&{1\over{2}}\, \sum_{\bf p} \Big( V^{+}_{\bf p}\,
\delta \bar{\rho}_{\bf- p}\,\delta \bar{\rho}_{\bf p} +
V^{-}_{\bf p}\,\delta \bar{d}_{\bf- p}\,\delta \bar{d}_{\bf p} \Big)
\nonumber\\ &&
+ \triangle H^{\rm Coul},
\end{eqnarray}
where
\begin{eqnarray}
V^{\pm}_{\bf p}={1\over{2}}\Big(V_{11}[{\bf p}] \pm 
V_{12}[{\bf p}] \Big).
\end{eqnarray}
With an  analogous $O(2)$ rotation
\begin{eqnarray}
A_{\mu}^{1}=A_{\mu}^{\rm em}+A_{\mu}^{-}, \
A_{\mu}^{2}=A_{\mu}^{\rm em}-A_{\mu}^{-},
\end{eqnarray}
$A_{\mu}^{\rm em}$ is coupled to  $\bar{\rho}$ and $A_{\mu}^{-}$ to
$\bar{d}$ in $\bar{H}$. 
Here $A_{\mu}^{\rm em}$ represents the electromagnetic potential that
probes in-phase density fluctuations of the two layers while
$A_{\mu}^{-}$ probes the out-of-phase density fluctuations.
The field-dependent Coulomb interaction term is now written as
$\triangle H^{\rm Coul}= \triangle^{+} H^{\rm Coul}
+\triangle^{-}H^{\rm Coul}$, with
\begin{eqnarray}
\triangle^{+} H^{\rm Coul}={1\over{2}}\sum_{\bf p} V^{+}_{\bf p}\,
&& \sum_{\bf k}\Big( u^{\rm em}[{\bf p,k}]\,
\{ \bar{\rho}_{\bf -p} , \bar{\rho}_{\bf p\! -\!k} \}
\nonumber\\ &&
+ u^{-}[{\bf p,k}]\,
\{ \bar{\rho}_{\bf -p} , \bar{d}_{\bf p\! -\!k}\, \} \Big),
\label{dplusH}
\end{eqnarray}
where $u^{\rm em}$ and $u^{-}$ stand for $u[{\bf p,k},t]$ in
Eq.~(\ref{uzerozero}) with $A_{\mu}$ replaced by $A_{\mu}^{\rm em}$
and $A_{\mu}^{-}$, respectively.
For $\triangle^{-} H^{\rm Coul}$ simply replace
$V^{+}[{\bf p}]
\rightarrow V^{-}[{\bf p}], \bar{\rho} \rightarrow \bar{d}$ and
$\bar{d} \rightarrow \bar{\rho}$ in the above.

We are now ready to discuss collective excitations within the lowest
Landau level. The projected single-mode approximation~\cite{GMP} (SMA)
provides a powerful means to study collective excitations of liquid
states and, in particular, shows the presence of gapful density
fluctuations in general single- and double-layer quantum Hall
systems.~\cite{RR,MZ,GMP}

Let $|G\rangle$ denote the exact ground state of the double-layer system
of our concern (for $A_{\mu}=0$).
The SMA supposes that the density fluctuations over $|G\rangle$ have
predominant overlap with $|G\rangle$ through the associated density
operators $\rho^{\alpha}$.  For the present system we consider two modes, 
a charge mode $|\phi^{+}_{\bf k}\rangle \sim \bar{\rho}_{\bf k}|G\rangle$ 
representing the in-phase density fluctuations of the two
layers  and a phonon-roton mode 
$|\phi^{-}_{\bf k}\rangle \sim \bar{d}_{\bf k}|G\rangle$
representing the interlayer out-of-phase density fluctuations.

Let us normalize the $|\phi^{-}_{\bf k}\rangle$ mode by writing
\begin{eqnarray}
|\phi^{-}_{\bf k}\rangle = {1\over{\sqrt{2N\, \bar{s}^{-}(k)}}}\,
\bar{d}_{\bf k}|G\rangle
\end{eqnarray}
where the normalization of the wave function
\begin{eqnarray}
\bar{s}^{-}({\bf k}) &\equiv& {1\over{2N}}\,
\langle G|\bar{d}_{\bf -k}\,\bar{d}_{\bf k}\, |G\rangle 
=\bar{s}^{-}({\bf -k})
\end{eqnarray}
is nothing but the (projected) static structure factor of the
ground state $|G\rangle$; $N=N^{1}+N^{2}$ denotes the total
number of electrons. [To be precise, $\bar{d}_{\bf k}$ in the above
refers to $\bar{d}_{\bf k}(t)$ at some fixed time, say,
$t=0$.]   
Similarly, we normalize $|\phi^{+}_{\bf k}\rangle$
with $\bar{s}^{+}({\bf k}) =\langle G|
\bar{\rho}_{\bf -k}\,\bar{\rho}_{\bf k}\, |G\rangle /(2N)$.

In the SMA the static structure factors $\bar{s}^{\pm}({\bf k})$
are the basic quantities, through which the effect of nontrivial
correlations pertinent to the ground state is reflected in
the dynamics.  Conservation laws reveal some of their general
features:~\cite{RR,MZ}
Invariance under translations of both layers implies an analog of
Kohn's theorem for the interlayer in-phase collective excitations so
that $\bar{s}^{+}({\bf k}) \sim |{\bf k}|^{4}$ for small 
${\bf k}$. As a result, the $O({\bf k})$ and $O({\bf k}^{2})$
in-phase ($A_{\mu}^{\rm em}$) response of the double-layer system is
governed by the cyclotron modes and is essentially the same as that of
the single-layer system.

On the other hand, the presence of interlayer interactions 
$V_{12}({\bf p})$ spoils invariance under relative translations of the
two layers and, unless spontaneous interlayer coherence is realized,
the out-of-phase collective excitations become
dipole-active,~\cite{RR,MZ}
\begin{equation}
\bar{s}^{-}({\bf k}) =
c^{-}\,{1\over{2}}\,{\bf k}^{2} +  O(|{\bf k}|^{4}).
\label{sminus}
\end{equation}
With this fact in mind we shall henceforth concentrate on the
dipole-active out-of-phase response. 
Actually,  particular sets of double-layer FQH states we have in mind are
the Halperin  $(m_{1},m_{2},n)$ or $(m,m,n)$  states.~\cite{BIH}
For the $(m,m,n)$ states the coefficient $c^{-}$ is explicitly
known:~\cite{RR}
\begin{eqnarray}
c^{-} = {n\over{m-n}}.
\label{cminus}
\end{eqnarray}
In this paper we do not discuss the case of the $(m,m,m)$ states,
where interlayer coherence develops~\cite{WZdlayer,EI}
and which thus requires a separate analysis.

In the SMA the excitation energy $\epsilon^{-}_{\bf k}$ of the
collective mode is determined from the (projected) oscillator strength
\begin{equation}
\bar{f}^{-}({\bf k}) = {1\over{2N}}\, \langle G|\,
\bar{d}_{\bf -k}\, (H - E_{0})\, \bar{d}_{\bf k}\, |G\rangle
\end{equation}
(with $E_{0}$ being the ground state energy) by saturating it with the
single mode
$|\phi^{-}_{\bf k}\rangle$, so that
\begin{equation}
\epsilon^{-}_{\bf k} = f^{-}({\bf k})/s^{-}({\bf k}).
\end{equation}
This $\bar{f}^{-}({\bf k})$ is directly calculated (by first rewriting
it as a double commutator
${1\over{4N}}\,  \langle G|\,
[\,\bar{d}_{\bf -k},\ [H\, ,\,\bar{d}_{\bf k}]\,]\,|G\rangle $ 
and) using the $W_{\infty}$ algebra of the projected charges,~\cite{GMP}
\begin{eqnarray}
&&[\bar{\rho}_{\bf p}, \bar{\rho}_{\bf k}]
=[\bar{d}_{\bf p}, \bar{d}_{\bf k}]
= (e^{{1\over{2}} p^{\dag}k} -e^{{1\over{2}} k^{\dag}p} )\, 
\bar{\rho}_{\bf p+k}, \nonumber\\
&&[\bar{\rho}_{\bf p}, \bar{d}_{\bf k}]
= (e^{{1\over{2}} p^{\dag}k} -e^{{1\over{2}} k^{\dag}p} )\,
\bar{d}_{\bf p+k},
\label{chargealgebra}
\end{eqnarray}
with the result~\cite{RR,MZ}
\begin{equation}
\bar{f}^{-}({\bf k})
= - {{\bf k}^{2}\over{2}} \sum_{\bf p}
{\bf p}^{2}\ V_{12}[{\bf p}]\, \bar{s}^{12} ({\bf p})
+ O(|{\bf k}|^{4}), 
\label{fminus}
\end{equation}
where $ \bar{s}^{12} ({\bf p}) \equiv 
\langle G|\bar{\rho}^{1}_{\bf -p}\,\bar{\rho}^{2}_{\bf p}\, |G\rangle /N
={1\over{2}}\, \{ \bar{s}^{+} ({\bf p}) -\bar{s}^{-} ({\bf p}) \}$. 
This leads to the SMA excitation gap at  ${\bf k}\rightarrow 0$:
\begin{equation}
\epsilon^{-}_{0}\equiv \epsilon^{-}_{\bf k=0}
={1\over{c^{-}}}\, \sum_{\bf p}V_{12}[{\bf p}]\,{\bf p}^{2}
\{-\bar{s}^{12}({\bf p})\},
\label{eminuszero}
\end{equation}
where $1/c^{-} =(m-n)/n$ for the $(m,m,n)$ states.

Let us now study the response of the FQH ground states, which
involves excitation of collective modes.
Consider first the $A^{-}_{0}$ to $A^{-}_{0}$ response, and write the
associated density-density response function 
$(-i)\langle G|T\bar{d}(x)\bar{d}(x') |G\rangle$ in spectral form in
Fourier space
\begin{eqnarray}
F[\omega,{\bf k}]
&=&\sum_{n}\Big\{{1\over{\omega  -\epsilon_{n}}}\, \sigma^{n}({\bf k})
-{1\over{\omega + \epsilon_{n}}}\, \sigma^{n}({\bf -k}) \Big\},
\label{spectralform}\\
\sigma^{n}({\bf k}) &=& {1\over{\Omega}}\,
\langle G|\, \bar{d}_{\bf k}\,  |n\rangle
\langle n|\,\bar{d}_{\bf -k} |G\rangle ,
\end{eqnarray}
where $\epsilon_{n}\equiv E_{n} -E_{0}$ stands for the excitation energy
of the intermediate state $|n\rangle$;
$\Omega$ denotes the spatial surface area.
In the SMA we saturate the sum over $|n\rangle$ by a single collective
mode $|\phi^{-}_{\bf k}\rangle$
so that 
$\sigma^{n}({\bf k}) \rightarrow (2N/\Omega)\,\bar{s}^{-}({\bf k})$,
and find, noting $\sigma^{n}({\bf k})=\sigma^{n}({\bf -k})$, 
\begin{equation}
F[\omega,{\bf k}]
=\rho_{\rm av}\,{4\,\epsilon^{-}_{\bf k}\over
{\omega^{2} -(\epsilon^{-}_{\bf k})^{2}}}\,\bar{s}^{-}({\bf k}) ,
\label{Fdd}
\end{equation}
where $\rho_{\rm av}=N/\Omega = \rho^{1}_{\rm av} + \rho^{2}_{\rm av}$.
With $\bar{s}^{-}({\bf k})\approx {1\over{2}}c^{-}{\bf k}^{2}$,
this leads to a dipole response of the form
\begin{equation}
S^{\rm col}_{A_{0}} =  {1\over{2}}\, \rho_{\rm av}
\int d^{3}x\,  \partial_{j}A^{-}_{0}(x) \,
{2\, c^{-}\,\epsilon^{-}_{0}\over{ 
(\epsilon^{-}_{0})^{2}} \!-\!\omega^{2}}\,
\partial_{j}A^{-}_{0}(x) ,
\label{AzeroAzero}
\end{equation}
where $\omega$ stands for $i\partial_{t}$.
In view of the simple structure of $A_{\mu}^{-}\bar{d}$ coupling in
$\bar{H}$ of Eq.~(\ref{HDL})  it may appear somewhat disturbing that
this response is not gauge invariant by itself. Fortunately, the
field-dependent Coulomb interaction $\triangle H^{\rm Coul}$ serves to
promote it into a gauge-invariant form and at the same time yields
another important response, the Chern-Simons (CS) term, as shown below.

Let us consider the $\triangle H^{\rm Coul}$ to $A_{0}^{-}$ response.
It contains correlation functions like
$\langle G| T\, d_{\bf k}(t)\, \bar{I}_{\bf p,k}(0)|G\rangle$,
which we write in  spectral form and saturate with a single collective
mode $|\phi^{-}_{\bf k}\rangle$ again, where
\begin{equation}
\bar{I}_{\bf p,k} =\{\bar{\rho}_{\bf -p},\bar{d}_{\bf p-k}\}.
\end{equation}
The result is
\begin{eqnarray}
 S^{\rm col}_{\triangle H}
= -\int dt \sum_{\bf k} A_{0}^{-}[{\bf -k}] \Big\{ 
&&
{1\over{\omega - \epsilon^{-}_{\bf k}}}\, \sigma_{d I}^{\phi}({\bf k})
\nonumber\\ &&
-{1\over{\omega + \epsilon^{-}_{\bf k}}}\,\sigma_{I d}^{\phi}({\bf -k})
\Big\},
\end{eqnarray}
with
\begin{eqnarray}
\sigma_{d I}^{\phi}({\bf k})
&=& {1\over{2\Omega}}\sum_{\bf p}  u^{-}[{\bf p,k}]
\Big\{ V^{+}_{\bf p}\, g({\bf p},{\bf k})
+ V^{-}_{\bf p}\,g({\bf k\!-\!p},{\bf k}) \Big\}, 
\nonumber\\
\label{sigmadI}\\
g({\bf p},{\bf k}) 
&=& \langle G|\, \bar{d}_{\bf k}\, \bar{I}_{\bf p,k} |G\rangle .
\end{eqnarray}
For $\sigma_{I d}^{\phi}({\bf -k})$ replace
$g({\bf p},{\bf k})$ in Eq.~(\ref{sigmadI}) by 
$\hat{g}({\bf p},{\bf k}) \equiv 
\langle G|\, \bar{I}_{\bf p,k}\,\bar{d}_{\bf k}\,|G\rangle$ 
which equals $g({\bf -p},{\bf -k})^{\dag}$.

It is now necessary to calculate the matrix elements of products
of three charges or the three-particle distribution in the ground state
$|G\rangle$.  A similar problem was discussed earlier by
Feenberg~\cite{Feenberg} in his attempt at improving the Bijl-Feynman
theory of the roton spectrum of liquid $^{4}$He.  His result adapted
to the present case reads 
\begin{equation}
\langle G| \bar{d}_{\bf k}\, \{\bar{\rho}_{\bf -p} ,
\bar{d}_{\bf p-k}\} |G\rangle \propto N \bar{s}^{-}({\bf k})\,
 \bar{s}^{+}({\bf p})\,  \bar{s}^{-}({\bf p-k})
\label{naiveproduct}
\end{equation}
provided the charges are mutually commuting like their unprojected
counterparts (or more precisely, the arguments ${\bf r}=(r_{1},r_{2})$ of
the projected charges  are taken to commute).
If Eq.~(\ref{naiveproduct}) were the case, $g({\bf p},{\bf k})$ in
$\sigma_{d I}^{\phi}({\bf k})$ would be of $O({\bf k}^{2})$, leading
to responses involving three derivatives or more, i.e., 
no $O({\bf k}^{2})$ response. 
This is the key observation and suggests that the desired 
$O({\bf k}^{2})$ response is determined  by isolating from the
products of three charges the portion that derives from the
noncommutative nature of the projected charges.  
Fortunately, for the $O({\bf k}^{2})$ response it suffices to
calculate $g({\bf p},{\bf k})$ to first order in ${\bf k}$; 
this is easily seen from the structure of $u^{-}[{\bf p,k}]$ 
[that the $p^{n}$ terms are of $O(k^{n-1})$] and by use of
symmetric integration over ${\bf p}$.

For actual calculation  we decompose the three-charge products into sums
of normal-ordered products with respect to ${\psi_{0}^{G}}^{\dag}$
and $\psi_{0}^{G}$, using the multiplication laws like
$\bar{\rho}_{\bf p}\,\bar{\rho}_{\bf k}
= e^{{1\over{2}}\, p^{\dag} k}\,\bar{\rho}_{\bf p+k} +
:\bar{\rho}_{\bf p}\,\bar{\rho}_{\bf k}: $.
The relevant noncommutative portion is thereby readily identified and
is seen to consist of products of two charges at most;
see Appendix A.
In particular, the noncommutative portion of 
$\bar{d}_{\bf k}\, \bar{I}_{\bf p,k}$ has the structure
\begin{eqnarray}
(\bar{d}_{\bf k}\bar{I}_{\bf p,k})^{\rm NC}&=&
 (i{\bf p\times k} + {\bf p\cdot k})\, 
(\bar{\rho}_{\bf -p}\bar{\rho}_{\bf p} 
-\bar{d}_{\bf k-p}\bar{d}_{\bf p-k} )
\nonumber\\ && 
+  O({\bf k}^{0})\, \bar{d}_{\bf -k}\bar{d}_{\bf k}
+ O({\bf k}^{2}),
\label{deltaIone}
\end{eqnarray}
which leads to
\begin{eqnarray}
\sigma_{d I}^{\phi}({\bf k})\!
&=& \!\rho_{\rm av} \sum_{\bf p}  u^{-}[{\bf p,k}]
\Big\{ 2 (i {\bf p\!\times \! k} \!+\!{\bf p\!\cdot \! k} ) V_{12}[{\bf p}]\, 
\bar{s}^{12}({\bf p}) + \cdots \Big\}, \nonumber\\
&=&  \rho_{\rm av} c^{-}\epsilon_{0}^{-}
(\delta^{jm}-i\epsilon^{jm} )\, k_{j} \Big\{A_{m}[{\bf k}] +
O(A_{i0})\Big\} %\nonumber\\&&\  
+ O({\bf k}^{3}) ,\nonumber\\
\label{sigmadH}
\end{eqnarray}
where we have used Eqs.~(\ref{uzerozero}) and~(\ref{eminuszero})
to arrive at the last line. 
Similarly, for $\sigma_{Id}^{\phi}({\bf -k})$ we obtain the same
expression~(\ref{sigmadH}) with $\delta^{jm} \rightarrow - \delta^{jm}$. 
[Here we remark that the difference
$\sigma_{d I}^{\phi}({\bf k}) - \sigma_{I d}^{\phi}({\bf -k})$ is
directly calculated by use of the charge algebra~(\ref{chargealgebra});
this offers an independent check of the calculation given above.]
The $O(A_{i0})$ terms in Eq.~(\ref{sigmadH}) give rise to nonleading
corrections to the dipole term~(\ref{AzeroAzero})  
[smaller by factor of $O(\epsilon_{0}^{-}/\omega_{c})$ or
$O(\partial_{t}/\omega_{c})$] 
and are omitted here.  Eventually one is led to a response of the form
\begin{eqnarray}
S^{\rm col}_{\triangle H}
&=&  \rho_{\rm av} \int d^{3}x\, A_{0}^{-}\,
{2 c^{-} \epsilon_{0}^{-}\,\over{(\epsilon^{-}_{0})^{2} -\omega^{2}}}\,
\Big[ \partial_{0}\partial_{j}A_{j}^{-}  
-\epsilon_{0}^{-}\epsilon^{ij}\partial_{i}A_{j}^{-} \Big] .
\nonumber\\
\label{SAzerou}
\end{eqnarray}

Likewise, the 
$\langle G|\,T \triangle H^{\rm Coul}\triangle H^{\rm Coul} |G\rangle$ 
response function gives rise to the
$\partial_{0}A^{-}_{j}(\cdots) \partial_{0}A^{-}_{j}$ and
$A^{-}_{i}(\cdots)\epsilon^{ij}\partial_{0}A^{-}_{j}$ terms that
precisely combine with Eqs.~(\ref{AzeroAzero}) and (\ref{SAzerou}) to
form gauge-invariant expressions, as shown in Appendix B.
Furthermore, substituting
$A^{-}_{0}\rightarrow A^{-}_{0} + (1/2M)\,A_{12}^{-}$ in
Eq.~(\ref{SAzerou}) generates a gauge-invariant
$A_{12}^{-}(\cdots) A_{12}^{-}$ term.

Finally, collecting terms so far obtained and adding also the
cyclotron-mode contribution in Eq.~(\ref{Scycl}) yields the complete
dipole and related responses in the SMA, 
\begin{eqnarray}
S_{\rm eff}^{-}
&=& {\rho_{\rm av}\over{2}}\int\! d^{3}x\, A^{-}_{j0}\, \Big[
{ 2\, c^{-}\,\epsilon^{-}_{0}\over{ (\epsilon^{-}_{0})^{2}}
\!-\! \omega^{2}}\, +
{\omega_{c}\over{ \omega _{c}^{2}} \!-\! \omega^{2}}\,\Big]
A^{-}_{j0} 
\nonumber\\  && 
- {\rho_{\rm av}\over{2}} \int\! d^{3}x\, A^{-}_{\mu} \,
\Big[{ 2 c^{-}\,(\epsilon_{0}^{-})^{2}
\over{(\epsilon^{-}_{0})^{2}-\omega^{2}}}
+{\omega_{c}^{2}\over{ \omega _{c}^{2}} \!-\!
\omega^{2}}\,\Big]\,\epsilon^{\mu \nu\rho}
\partial_{\nu}A^{-}_{\rho} 
\nonumber\\ && 
-{\rho_{\rm av}\over{2M}} \int\! d^{3}x\, A_{12}^{-}\,
\Big[\,
{ 2\, c^{-}\,(\epsilon^{-}_{0})^{2}
\over{(\epsilon^{-}_{0})^{2}} \!-\! \omega^{2}}\, +
{\omega_{c}^{2}\over{ \omega _{c}^{2}} \!-\! \omega^{2}}\,\Big]
 A_{12}^{-}.
\label{fullresponse}
\end{eqnarray}
In the  low-energy $(\omega \ll \epsilon^{-}_{0})$ regime
$S_{\rm eff}^{-}$ reads
\begin{eqnarray}
S_{\rm eff}^{-}
&\approx& {1\over{2}}\, \rho_{\rm av}\int d^{3}x\, 
\Big[ { 2\, c^{-}\over{ \epsilon^{-}_{0}}}(A^{-}_{j0})^{2}
\nonumber\\ &&
-( 2 c^{-}\!+1) \Big\{A^{-}_{\mu} \,\epsilon^{\mu \nu\rho}
\partial_{\nu}A^{-}_{\rho}
+{1\over{M}}\,(A_{12}^{-})^{2}\Big\}\Big].
\end{eqnarray}
The effect of the collective mode is significant here.
The dipole response acquires the scale $\sim
\epsilon^{-}_{0}/(2c^{-})$, and the Hall conductance and magnetic
susceptibility are enhanced by common factor $(2 c^{-}+1)$, 
which equals $(m+n)/(m-n)$ for the $(m,m,n)$ states. 
Here we see that the static structure factor of the $(m,m,n)$ states
precisely reproduces the effect of statistical-flux attachment employed 
in the CS theories: The Hall conductance
$\sigma^{-}_{xy}=-e^{2}\rho_{\rm av}\,(m+n)/(m-n)$ is
exactly what one gets in the CS theories~\cite{EI,LFdlayer},
where, however, the collective-excitation spectrum is put on the scale
of $\omega_{c}$.

Note here that the static density correlation functions are not
sensitive to such a scale change.~\cite{MZ} 
They are given by
$\rho _{\rm av} (2c^{-}+1)\, {\bf p}^{2}={\bf p}^{2}/\{\pi (m-n)\}$ 
for $\langle G| d_{\bf -p}d_{\bf p}|G\rangle$ and 
$\rho _{\rm av}\, {\bf p}^{2}={\bf p}^{2}/\{\pi (m+n)\}$ for
$\langle G| \rho_{\bf -p}\rho_{\bf p}|G\rangle$ in the long wavelength
limit.
In this way, the cyclotron mode and the collective mode contribute to 
the (unprojected) static structure factor in comparable magnitude,
i.e., 1 to $2c^{-}$ in ratio so that for the oscillator strength the
collective-mode contribution is smaller by factor 
$2c^{-}(\epsilon^{-}_{0}/\omega_{c}) < 1$.

It will perhaps be worth remarking here that, while the SMA is
a variational method in principle, it has led to an
electromagnetic response of gauge-invariant form.
This is consistent with an earlier observation~\cite{GMP} 
that the SMA is better suited for quantum Hall systems than the case of
helium where the backflow corrections were necessary for restoring
gauge invariance.

\section{effective gauge theory}
 
In the previous section we have calculated the response of a
double-layer electron system. Once such a response is known, it is
possible to reconstruct it through the quantum fluctuations of a boson
field. This procedure, known as functional bosonization,~\cite{FS}
was previously applied to single-layer systems to derive~\cite{KSbos}
an effective theory of the FQHE. In this section we construct an
effective theory for the double-layer system.

For bosonization we take a shortcut and quote only a general formula.
Consider a three-vector field $b_{\mu}$ coupled to an external field
$A_{\mu}$, with a Lagrangian of the form
\begin{eqnarray}
L[b]&=&  -A_{\mu} \epsilon^{\mu \nu\rho}
\partial_{\nu}b_{\rho} + {1\over{2}}\, b_{\mu}
\epsilon^{\mu\nu\rho}\,\beta\,\partial_{\nu} b_{\rho}
\nonumber\\  &&
+{1\over{2}}\,b_{k0}{\beta\over{\alpha}}\,b_{k0}
-{1\over{2}}\, b_{12}\,\beta\,\sigma\,b_{12} -\kappa\, b_{0},
\label{LbA}
\end{eqnarray}
where
$\beta$, $\alpha$ and $\sigma$ may contain derivatives and
$\kappa$ is a  real constant;
$b_{\mu\nu}\equiv \partial_{\mu}b_{\nu} -\partial_{\nu}b_{\mu}$
and $\epsilon^{012} =1$.
Direct functional integration over $b_{\mu}$ (with suitable gauge
fixing) shows that this vector-field theory leads to an effective 
Lagrangian of the form~\cite{KSbos} 
\begin{eqnarray}
 L_{\rm eff}[A]&=& 
- {1\over{2}}\,A_{\mu}{\alpha^{2}\over{\beta \cal D}}\,
\epsilon^{\mu \nu\rho} \partial_{\nu} A_{\rho}
+{1\over{2}}\, A_{k0}{\alpha \over{\beta \cal D}} A_{k0}
\nonumber\\ &&
- {1\over{2}}\,A_{12} {\sigma \alpha^{2}\over{\beta \cal D}} A_{12}
-{\kappa\over{\beta}}\,A_{0}
\label{resultL}
\end{eqnarray}
in obvious compact notation, where
\begin{eqnarray}
&&{\cal D}=\alpha^{2} +\partial_{t}^{2} - \alpha\,\sigma\,\nabla^{2} .
\end{eqnarray}

With this formula, it is straightforward to verify that, of the
out-of-phase response $S_{\rm eff}^{-}$ in Eq.~(\ref{fullresponse}), 
the collective-mode contribution is reconstructed from the theory of a
vector field $\xi_{\mu}$, with the Lagrangian
\begin{eqnarray}
L^{\rm col}_{\xi}&=& -A_{\mu}^{-} \epsilon^{\mu \nu\rho}
\partial_{\nu}\xi_{\rho}
\nonumber\\  &&
+ {1\over{4 c^{-}\rho_{\rm av}}}
\Big[ \xi_{\mu} \epsilon^{\mu\nu\rho}\partial_{\nu} \xi_{\rho}
+ {1\over{\epsilon_{0}^{-}}}\,(\xi_{k0})^{2}
- {1\over{M}}\, (\xi_{12})^{2}\Big].
\label{Lxi}
\end{eqnarray}
Similarly, the  cyclotron-mode contribution $S^{\rm cycl}$ of
Eq.~(\ref{Scycl}), to the dipole and related order, is reproduced from
an analogous Lagrangian consisting of a pair of vector fields
$b^{\alpha}_{\mu}=(b^{1}_{\mu},b^{2}_{\mu})$,
\begin{eqnarray}
L_{b}^{\rm cycl}
&=&-A^{\alpha} \epsilon \partial b^{\alpha}
- b^{\alpha}_{0} 
\nonumber\\  &&
+ {1\over{2\rho^{\alpha}_{\rm av}}}
\Big[ b^{\alpha} \epsilon \partial b^{\alpha}
+{1\over{\omega_{c}}}\,(b^{\alpha}_{k0})^{2} 
- {1\over{M}}\,(b^{\alpha}_{12})^{2}\Big] ,
\label{Lbminus}
\end{eqnarray}
where summations over layer indices $\alpha$ are understood;
$A^{\alpha} \epsilon \partial b^{\alpha} \equiv A_{\mu}^{\alpha}
\epsilon^{\mu\nu\rho} \partial_{\nu} b_{\rho}^{\alpha}$, etc., for short.
Thus the double-layer system is now described by a gauge-field theory with
the simple Lagrangian $L^{\rm col}_{\xi}+L_{b}^{\rm cycl}$.

The Lagrangian $L_{b}^{\rm cycl}$ reminds us of the dual-field Lagrangian
of Lee and Zhang~\cite{LZ} (LZ), derived within the
Chern-Simons-Landau-Ginzburg theory and describing the long-wavelength 
characteristics of the FQHE. 
The LZ Lagrangian, generalized to describe 
the $(m_{1},m_{2},n)$ Halperin states of the double-layer system, 
is written in terms of a pair of vector fields
$b^{\alpha}_{\mu}=(b^{1}_{\mu},b^{2}_{\mu})$:
\begin{eqnarray}
L^{\rm CS}_{\rm eff}[b]
&=& - A^{\alpha} \epsilon\partial b^{\alpha} - b^{\alpha}_{0} 
+ {1\over{2}}\,\Lambda_{\alpha \beta}\,b^{\alpha} \epsilon
\partial b^{\beta}
\nonumber\\ &&
+{1\over{2}}\,{1\over{\omega_{c}\rho^{\alpha}_{\rm av}}}\,
(b^{\alpha}_{k0})^{2} + \cdots.
\label{LCSeff}
\end{eqnarray}
The omitted terms $(\cdots)$ contain Coulomb interactions and higher
derivative terms, not relevant to our present discussion.
The mixing matrix  
\begin{eqnarray}
\Lambda_{\alpha \beta}&=&
2\pi \left(\begin{array}{cc}
m_{1} & n\\
  n   & m_{2}
           \end{array} \right)
\label{LCSab}
\end{eqnarray}
is a consequence of statistical-flux attachment characteristic of the
$(m_{1},m_{2},n)$ states.

The LZ Lagrangian~(\ref{LCSeff}) rests on the composite-boson picture
of the FQHE.  Actually it is possible~\cite{KSbos} to derive it as an
effective theory reconstructed from the response of the composite 
fermions.   
Thus $L^{\rm CS}_{\rm eff}[b]$ is a consequence of the CS theories, both
bosonic and fermionic.~\cite{fnLCS}

Our effective Lagrangian $L^{\rm col}_{\xi}+L_{b}^{\rm cycl}$
consists of three vector fields $(\xi_{\mu}, b_{\mu}^{\alpha})$
representing one collective mode and two cyclotron modes.
To make a connection with $L^{\rm CS}_{\rm eff}[b]$ let us try to
rewrite  
$L_{\xi}^{\rm coll}+ L_{b}^{\rm cycl}$ in favor of the new fields
$\eta_{\mu}^{1}=b_{\mu}^{1} +{1\over{2}}\xi_{\mu}$ and
$\eta_{\mu}^{2}=b_{\mu}^{2} -{1\over{2}}\xi_{\mu}$, eliminate
the collective mode $\xi_{\mu}$ and retain terms up to $O(\nabla^{2})$.
The result is
\begin{eqnarray}
L_{\rm eff}[\eta]
&=& - A^{\alpha} \epsilon\partial \eta^{\alpha} - \eta^{\alpha}_{0} 
+ {1\over{2}}\,\hat{\Lambda}_{\alpha\beta}\,\eta^{\alpha}
\epsilon \partial \eta^{\beta}
\nonumber\\  &&
+{1\over{2}}\,\Big[{1\over{\omega_{c}}}\,\hat{\Lambda}_{\alpha\beta} 
+ ({1\over{\epsilon_{0}^{-}}} - {1\over{\omega_{c}}})\Xi_{\alpha\beta}
\Big] \eta^{\alpha}_{k0}\eta^{\beta}_{k0}
\nonumber\\ &&
-{1\over{2M}}\,\hat{\Lambda}_{\alpha \beta}
\eta^{\alpha}_{12}\eta^{b}_{12} + \cdots,
\end{eqnarray}
with the mixing matrices
\begin{eqnarray}
\hat{\Lambda}_{\alpha \beta}&=& 2\pi \lambda
 \left(\begin{array}{cc}
1+ (2/c^{-})\nu_{2}/\nu  &1\\
  1 & 1+ (2/c^{-})\nu_{1}/\nu
      \end{array} \right) ,
\label{Lab}\\
\Xi_{\alpha \beta}&=&  2\pi \lambda_{\Xi}
\left(\begin{array}{cc}
\nu_{2}/\nu_{1}& -1\\
  -1 &  \nu_{1}/\nu_{2}
      \end{array} \right), \\
\lambda &=&1/\Big(\nu +{2\over{c^{-}}}{\nu_{1}\nu_{2}\over{\nu}}
\Big), \ \ \ 
\lambda_{\Xi}={2\over{c^{-}}}
{\nu_{1}\nu_{2}\over{\nu}}\,\lambda^{2},
\end{eqnarray}
where $\nu_{\alpha}=2\pi\ell^{2}\rho_{\rm av}^{\alpha}$ denote the
filling factors for each layer and $\nu = \nu_{1}+\nu_{2}$.
Note that $\Xi_{\alpha \beta}$ is essentially a projection operator.

Let us here take as the ground state $|G\rangle$ the $(m_{1},m_{2},n)$
state, for which $\nu_{1}=(m_{2}-n)/(m_{1}m_{2}-n^{2})$ and
$\nu_{2}=(m_{1}-n)/(m_{1}m_{2}-n^{2})$.
Try to ask if $L_{\rm eff}[\eta]$ and $L^{\rm CS}_{\rm eff}[b]$ could
share the same $O(\partial)$ long-wavelength structure or the CS
term, i.e.~, $\hat{\Lambda}_{\alpha \beta}=\Lambda_{\alpha \beta}$.
The answer is affirmative. This fixes $c^{-}$ uniquely,
\begin{equation}
c^{-}={2n\over{m_{1}+m_{2}- 2n}},
\label{cminustwo}
\end{equation}
and yields $\lambda=n$ and $\lambda_{\Xi}=n(1-n\nu)$.  We have thus
determined what appears to be the static structure factor to 
$O({\bf k}^{2})$ of the $(m_{1},m_{2},n)$ state, though not calculated
directly so far; it is correctly reduced to the known value~(\ref{cminus})
for the $(m,m,n)$ state.

While $L_{\rm eff}[\eta]$ and $L^{\rm CS}_{\rm eff}[b]$ coincide to
$O(\partial)$, they differ in $O(\partial^{2})$,
especially in the out-of-phase collective-mode spectrum.
Let us explore the difference in detail for the $(m,m,n)$ states.
The $L_{\rm eff}[\eta]$ and $L^{\rm CS}_{\rm eff}[b]$ are split into
two sectors with
$\eta_{\mu}^{\pm}=\eta_{\mu}^{1} \pm \eta_{\mu}^{2}$ and 
$b_{\mu}^{\pm}=b_{\mu}^{1} \pm b_{\mu}^{2}$, respectively.  
The $\eta_{\mu}^{+}$ sector is given by the Lagrangian~(\ref{LbA}) with
$b_{\mu} \rightarrow \eta_{\mu}^{+},
A_{\mu} \rightarrow A_{\mu}^{\rm em}, \kappa \rightarrow 1$,
$\beta \rightarrow 1/\rho_{\rm av}= \pi (m+n)$ and 
$\alpha \rightarrow \omega_{c}$, in precise agreement with the
$b_{\mu}^{+}$ sector of $L^{\rm CS}_{\rm eff}[b]$.  
On the other hand, the $\eta^{-}_{\mu}$ sector is given by the
Lagrangian~(\ref{LbA}) with
$b_{\mu} \rightarrow \eta_{\mu}^{-}, A_{\mu} \rightarrow A^{-}_{\mu}$,
$\kappa \rightarrow 0$ and
\begin{equation}
\beta  \rightarrow  \pi (m-n),\ \ \
{1\over{\alpha}} \rightarrow {1\over{\epsilon_{0}^{-}}}\,
{2n\over{m+n}} +{1\over{\omega_{c}}}\,{m-n\over{m+n}} .
\label{bainv}
\end{equation}
The $b_{\mu}^{-}$ sector of $L^{\rm CS}_{\rm eff}[b]$ differs merely
by the collective-mode spectrum 
\begin{equation}
\alpha\rightarrow 
\alpha^{-}_{\rm CS}=\pi (m-n)\,\rho_{\rm av}\,\omega_{c}
={m-n\over{m+n}}\,\omega_{c}.
\label{alphaCS}
\end{equation}
The static density correlation functions $\sim (1/\beta) {\bf p}^{2}$
are read from these $\beta$ values $\pi (m\pm n)$ and agree in both
theories, as noted in Eq.~(\ref{fullresponse}).
The $1/\alpha$ in Eq.~(\ref{bainv}) correctly recovers the energy
scale in the $\omega\rightarrow 0$ limit of Eq.~(\ref{fullresponse});
here the collective-mode energy is apparently shifted to
${m+n\over{2n}}\,\epsilon^{-}_{0}$ from the on-resonance value
$\epsilon^{-}_{0}$.

From the above consideration emerge the following observations:
The presence of the dipole-active out-of-phase collective excitations,
inherent to double-layer systems in general and specifically to the
$(m_{1},m_{2},n)$ Halperin states, implies strong interlayer
correlations that affect the $O(\partial)$ and $O(\partial^{2})$
long-wavelength characteristics of the double-layer FQH states.
The leading $O(\partial)$ correlations are correctly incorporated
into the CS theories by the flux attachment transformation, 
which, however, fails to take in the next-leading $O(\partial^{2})$
correlations at least in the random-phase approximation; 
as a result, the collective-excitation spectrum is left on the scale
of $O(\omega_{c})$.  In view of this, a practical way to improve
the CS theories would be to regard the collective-excitation energy
as a parameter to be adjusted phenomenologically.  

Finally we wish to discuss vortex excitations in double-layer
systems. An incompressible FQH state supports vortex excitations.~\cite{L} 
Vortices are readily introduced into the bosonic effective theories by
replacing $A^{\alpha}_{\mu} \rightarrow
A^{\alpha}_{\mu}+\partial_{\mu}\Theta^{\alpha}$, where
$\Theta^{\alpha} (x)$ stands for a topologically nontrivial
component of the phase of the electron field in each layer.~\cite{LZ}
The $L^{\rm CS}_{\rm eff}[b]$ and $L_{\rm eff}[\eta]$ thereby acquire 
a vortex coupling of the same form
$-2\pi\,\tilde{j}^{\alpha}_{\mu}b^{\alpha}_{\mu}$ and
$-2\pi\,\tilde{j}^{\alpha}_{\mu}\eta^{\alpha}_{\mu}$, respectively,
where $\tilde{j}^{\alpha}_{\mu}=
(1/2\pi)\,\epsilon^{\mu\nu\rho}\partial_{\nu}\partial_{\rho}
\Theta^{\alpha}$ denotes the vortex three-current
\begin{equation}
\tilde{j}^{\alpha}_{\mu}(x)= \sum_{i}
[1,\partial_{t}{\bf x}^{(i)}(t) ]\,
q^{\alpha}_{i}\,\delta^{2} ({\bf x} -{\bf x}^{(i)}(t)) ,
\end{equation}
with ${\bf x}^{(i)}(t)$ standing for the trajectory of the $i$th vortex
of vorticity $q^{\alpha}_{i}=\pm1,\pm2,\cdots$ in layer $\alpha$.
The vortex charges are easily read from these vortex
couplings.~\cite{KSbos} 
The electromagnetic coupling 
$- A^{\alpha}\epsilon\partial b^{\alpha}$ induces in
$b^{\alpha}_{\mu}$ an $A_{\mu}^{\alpha}$-dependent piece which is
isolated by writing 
$b^{\alpha}_{\mu}={b'}^{\alpha}_{\mu} +f^{\alpha}_{\mu}[A]$ and
choosing
$f^{\alpha}_{\mu}[A]=(\Lambda^{-1})_{\alpha\beta}A^{\beta}_{\mu} 
+ O(\partial A)$.
As a result, the vortex is coupled to $A^{\alpha}_{\mu}$ through
this background piece
$-2\pi\,\tilde{j}^{\alpha}_{\mu}b^{\alpha}_{\mu}=
-2\pi\,\tilde{j}^{\alpha}_{\mu}(\Lambda^{-1})_{\alpha\beta}
A^{\beta}_{\mu} +\cdots$. 
This reveals that a vortex of vorticity $(q^{1},q^{2})$ in
each layer induces the amount of charge
$Q^{\alpha}_{\rm v}=-2\pi\,q^{\beta}_{i}\, (\Lambda^{-1})_{\beta\alpha}$
in the two layers, or explicitly,
\begin{eqnarray}
Q^{1}_{\rm v}
=-e\,{m_{2}q^{1} -n q^{2}\over{m_{1}m_{2}- n^{2}}},\ 
Q^{2}_{\rm v}
=-e\,{-nq^{1} +m_{1} q^{2}\over{m_{1}m_{2}- n^{2}}}
\label{Qquasi}
\end{eqnarray}
for the $(m_{1}m_{2},n)$ states, in agreement with earlier
results.~\cite{EI}

\section{Summary and discussion}

In this paper we have studied the electromagnetic characteristics of
double-layer quantum Hall systems and derived an effective gauge
theory of the FQHE. 
Our approach has clarified, above all, that it is the dipole-active
excitations, both elementary and collective, that govern the transport
properties of quantum Hall systems. 
In particular, single-layer systems support no dipole-active
intra-Landau-level collective excitations, and consequently the
incompressible FQH states show universal long-wavelength
electromagnetic characteristics, governed by the inter-level cyclotron
mode alone. In contrast, for double-layer systems interlayer
out-of-phase collective excitations become dipole-active and alter the
response of the systems fundamentally.  
The effective theory constructed from the response via bosonization 
is written in terms of three vector fields which precisely reflect 
the three dipole-active modes, i.e., one out-of-phase collective mode
and two cyclotron modes, and properly incorporates the
single-mode-approximation spectrum of collective excitations.

The presence of the dipole-active interlayer collective excitations
implies strong interlayer correlations that affect the $O(\partial)$
and $O(\partial^{2})$ long-wavelength characteristics of the
double-layer FQH states.
The $O(\partial)$ correlations are correctly taken care of by
the flux attachment transformation in the Chern-Simons theories, both
bosonic and fermionic. The flux attachment, however, fails to take in
the next-leading $O(\partial^{2})$ correlations (at least in the
random-phase approximation). 
This explains why the Chern-Simons theories properly account for
$O(\partial)$ features, like the Hall conductance, vortex charges and
long-range orders, while they leave the collective-excitation spectrum
on the scale of the Landau gap $\sim \omega_{c}$.

The bosonization approach presented here gives rise to an effective
theory of the FQHE in a manner logically independent of the
Chern-Simons theories. It by itself does not tell at which filling
fractions the FQH states emerge. Instead, it tells us that the
long-wavelength characteristics of the incompressible FQH states are
determined independent of the composite-boson or composite-fermion
picture. It has thus allowed one to derive, for double-layer systems,
an effective theory that embodies the SMA spectrum of collective
excitations.

\acknowledgments

This work is supported in part by a Grant-in-Aid for Scientific
Research from the Ministry of Education of Japan, Science and Culture
(No. 10640265). 

\appendix

\section{Reduction of three-charge products}
In this appendix we outline how to isolate from products of
projected charges the portion that originates from the
noncommutative nature of $r_{i}$.
Consider a product of the form
$I= \bar{d}_{{\bf p}_{1}}\bar{\rho}_{{\bf p}_{2}}\,
\bar{d}_{{\bf p}_{3}}$.
By repeated use of the multiplication laws like 
$\bar{\rho}_{\bf p}\,\bar{d}_{\bf k}
= e^{{1\over{2}}\, p^{\dag} k}\,\bar{d}_{\bf p+k} 
+ :\bar{\rho}_{\bf p}\,\bar{d}_{\bf k}: $
one can decompose $I$ into normal-ordered products:
\begin{eqnarray}
\bar{d}_{{\bf p}_{1}}\bar{\rho}_{{\bf p}_{2}}\,\bar{d}_{{\bf p}_{3}}
&=& e^{{1\over{2}} [p^{\dag}_{1}(p_{2}+p_{3})+p^{\dag}_{2}p_{3}]}\,
\bar{\rho}_{{\bf p}_{1}+{\bf p}_{2}+{\bf p}_{3}}
\nonumber\\ && 
+ e^{{1\over{2}} p^{\dag}_{2}p_{3}}
:\bar{d}_{{\bf p}_{1}}\bar{d}_{{\bf p}_{2}+{\bf p}_{3}}:
+ e^{{1\over{2}} p^{\dag}_{1}p_{3}}
:\bar{\rho}_{{\bf p}_{2}}\bar{\rho}_{{\bf p}_{1}+{\bf p}_{3}}:
\nonumber\\ &&
+e^{{1\over{2}} p^{\dag}_{1}p_{2}}
:\bar{d}_{{\bf p}_{1}+{\bf p}_{2}}\bar{d}_{{\bf p}_{3}}:
+ :\bar{d}_{{\bf p}_{1}}\bar{\rho}_{{\bf p}_{2}}\,
\bar{d}_{{\bf p}_{3}}:.
\label{drhod}
\end{eqnarray}
The momentum dependent coefficients  all derive from the
operator nature of $r_{i}$. Thus it is easy to identify
the intrinsically noncommutative contribution 
\begin{eqnarray}
I^{\rm NC}
&=& (e^{{1\over{2}} p^{\dag}_{2}p_{3}}-1)\,
\bar{d}_{{\bf p}_{1}} \bar{d}_{{\bf p}_{2}+{\bf p}_{3}}
+(e^{{1\over{2}} p^{\dag}_{1}p_{3}}-1)\,
\bar{\rho}_{{\bf p}_{2}}\bar{\rho}_{{\bf p}_{1}+{\bf p}_{3}}
\nonumber\\ &&
+(e^{{1\over{2}} p^{\dag}_{1}p_{2}}-1)\,
\bar{d}_{{\bf p}_{1}+{\bf p}_{2}}\bar{d}_{{\bf p}_{3}}
+C\,\bar{\rho}_{{\bf p}_{1}+{\bf p}_{2}+{\bf p}_{3}} ,
\end{eqnarray}
where
\begin{eqnarray}
C &=& e^{{1\over{2}} p^{\dag}_{1}(p_{2}+p_{3})}-1
-(e^{{1\over{2}} p^{\dag}_{1}p_{3}}-1)
e^{{1\over{2}} p^{\dag}_{2}(p_{1}+p_{3})}
\nonumber\\ &&
- (e^{{1\over{2}} p^{\dag}_{1}p_{2}}-1)
e^{{1\over{2}} (p^{\dag}_{1}+p^{\dag}_{2}) p_{3}}.
\end{eqnarray}
It is now clear how to write down analogous formulas
for general three-charge products.

\section{Calculation of a response}

In this appendix we outline the calculation of the correlation
function  
$-i\langle G|T \triangle H^{\rm Coul}\triangle H^{\rm Coul} |G\rangle$
contributing to $S_{\rm eff}^{-}$ in Eq.~(\ref{fullresponse}).
Let us write it in spectral form and saturate it with a single
collective mode $|\phi^{-}_{\bf k}\rangle$.  The spectral weight
thereby takes a simple and suggestive form 
\begin{equation}
\sigma_{\triangle H \triangle H}^{\phi}({\bf k})
= \Omega\,  {1\over{2N\bar{s}^{-}({\bf k})}}
\sigma_{I d}^{\phi}({\bf -k})\sigma_{d I}^{\phi}({\bf -k}),
\end{equation}
with $\sigma_{d I}^{\phi}({\bf k})$ and
$\sigma_{I d}^{\phi}({\bf k})$ defined in Eq.~(\ref{sigmadI}).
On substitution of Eq.~(\ref{sigmadH}) and removing total divergences, 
$\sigma_{I d}^{\phi}({\bf -k}) \sigma_{d I}^{\phi}({\bf -k})$  turns
into $\gamma\, A_{i}[{\bf -k}] {\bf k}^{2} A_{i}[{\bf k}]$  
with $\gamma = (\rho _{\rm av} c^{-} \epsilon_{0}^{-})^{2}$ 
while
$\sigma_{I d}^{\phi}({\bf -k})\omega \sigma_{d I}^{\phi}({\bf -k})$
yields 
$-i\gamma\,\epsilon^{ij}A_{i}[{\bf -k}]
\omega {\bf k}^{2}A_{j}[{\bf k}]$.
These give rise to the
$\partial_{0}A^{-}_{j}[(\epsilon^{-}_{0})^{2}-\omega^{2}]^{-1}
\partial_{0}A^{-}_{j}$ and
$\epsilon^{ij}A_{i}[(\epsilon^{-}_{0})^{2}-\omega^{2}]^{-1}
\partial_{0}A^{-}_{j}$ terms in Eq.~(\ref{fullresponse}).

%%%%%%%%%%%%%%%%%%%%% References %%%%%%%%%%%%%%(trimmed)%%%%%%%%%


\begin{thebibliography}{99}
\bibitem{fnutwo} 
The $\partial_{k}A_{12}$ terms in the first term of ${\cal U}_{2}(x)$
can be removed  by replacing $D'\rightarrow D'+{1\over{2}}\nabla^{2}D$
in the second term and $D'\rightarrow D' +{1\over{4}}\nabla^{2}D$ 
in the last term.
\bibitem{fnLCS} 
To be precise, the $(b^{\alpha}_{12})^{2}$ terms turn out
picture-dependent. The composite-boson theory gives a vanishing
coefficient for it while the composite-fermion theory yields
coefficient $1/(2M\rho^{\alpha}_{\rm av})$.  
Thus the bosonic and fermionic Chern-Simons theories start to differ
at shorter wavelengths.
\end{thebibliography}
\end{document}